\documentclass[11pt,preprint]{aastex}

\def\giant{J1004+411}
\def\pg{PG1115+080}
\def\kpc{{\rm\,kpc}}
\def\shear{\phi_\gamma}
\def\grad{\phi_{\scriptscriptstyle\nabla}}
\def\radinx{\alpha}
\def\knbr{\langle\kappa_{\rm nbr}\rangle}
\def\oguri{O4}
\def\hoverarrow#1{\setbox0\hbox to 0pt{\hss$\scriptstyle\rightarrow$}%
                 #1\kern.6ex\raise 1.6ex\box0\kern-.2ex}
\def\btheta{{\hoverarrow\theta}}
\def\bthetan{{\hoverarrow\theta\!\!_n}}

\begin{document}

\title{Models of the giant quadruple quasar SDSS~J1004+4112}

       \author{Liliya L.R. Williams}
       \affil{Department of Astronomy\\
              University of Minnesota\\
	      116 Church Street SE\\
              Minneapolis, MN 55455}
       \email{llrw@astro.umn.edu}

       \and

       \author{Prasenjit Saha}
       \affil{Astronomy Unit\\
              Queen Mary and Westfield College\\
              University of London\\
              London E1~4NS, UK}
       \email{p.saha@qmul.ac.uk}

\begin{abstract}
SDSS~J1004+4112 is an unprecedented object.  It looks much like
several quadruple quasars lensed by individual galaxies, only it is
$\sim10$ times larger, and the lens is a cluster dominated by dark
matter.  We present free-form reconstructions of the lens using
recently-developed methods. The projected cluster mass profile is
consistent with being shallow, $r^{-0.3...-0.5}$, and can be fit 
with either an NFW or a flat-cored 3 dimensional mass distribution.  
However, we cannot rule out projected profiles as steep as $r^{-1.3}$.
The projected mass 
within $100\kpc$ is well-constrained as $(5\pm1)\times10^{13}M_\odot$, 
consistent with previous simpler models. Unlike previous work, however, 
we are able to detect structures in the lens associated with cluster 
galaxies. We estimate the mass associated with these galaxies, and
show that they contribute not more than about 10\% of the total cluster 
mass within $100\kpc$. Typical galaxy masses, combined with typical 
luminosities yield a rough estimate of their mass-to-light ratio,
which is in the single digits.
Finally, we discuss implications for time-delay measurements in this 
system, and possibilities for a partial Einstein ring.
\end{abstract}


\section{Introduction}

At the present time about 30 quadruply imaged QSO systems are known
\citep{koch04}. The majority of these are lensed by individual galaxies,
sometimes aided by a nearby group or cluster. Given the typical masses
of galaxies one expects that the typical image separation would be  
around 1 second of arc. This is borne out in the data. The exceptions,
i.e. large separation lenses are rare, and have to be produced by 
groups or clusters, not isolated galaxies. In fact, the very first
multiply imaged system, Q0957+561A,B \citep{walsh79} is a cluster lens 
with two of its images separated by $\sim 6''$. The recently discovered
SDSS~J1004+4112 (hereafter \giant) is the first system where a galaxy
cluster, $z=0.68$, splits a background QSO, $z=1.734$, into four images, 
producing an unprecedented image separation of $\sim 14''$ \citep{inada03}.
While this is the first such system to be discovered, Oguri et al. (2004)
estimate that the full Sloan Digital Sky Survey will yield several
similar large separation QSO lenses.

The image configuration of \giant~is almost identical to the
first-known quad \pg\ (if we swap NE and SW) but the scale is
$\sim10$ times larger, hence our description of \giant\ as a giant
quadruple quasar. The giantness prompts us to ask: is this cluster
lens just a scaled-up version of a typical galaxy lens, or is it
fundamentally different?  We will answer this question by constructing
and studying ensembles of pixelated mass maps for \giant\ and comparing
them with analogous mass maps for \pg.

\cite{oguri04} have modeled the system in various configurations of 
a main galaxy, cluster halo, and external shear. Even though a wide
range of mass distributions is consistent with the lensing data, a few
general conclusions apply to most mass models. For example, the mass
in the vicinity of the images is elongated North-South, consistent
with the distribution of visible galaxies. On larger scales, i.e.
well outside the image ring, the cluster's mass distribution must be
such that it gives rise to a large shear. \oguri~also conclude
that the cluster's center must be offset from that of the brightest 
galaxy by several kiloparsec. 
In the rest of the paper we model the \giant~cluster-lens in detail.
We agree with most of the general conclusions reached by \oguri,
even though our modeling method is very different from theirs. 
Unlike the parametric modeling in \oguri, our modeling allows us to
detect the mass inhomogeneities in the cluster associated with the
individual galaxies.


\section{Modeling the cluster}\label{modeling}

\begin{figure}
\epsscale{0.5}
\plotone{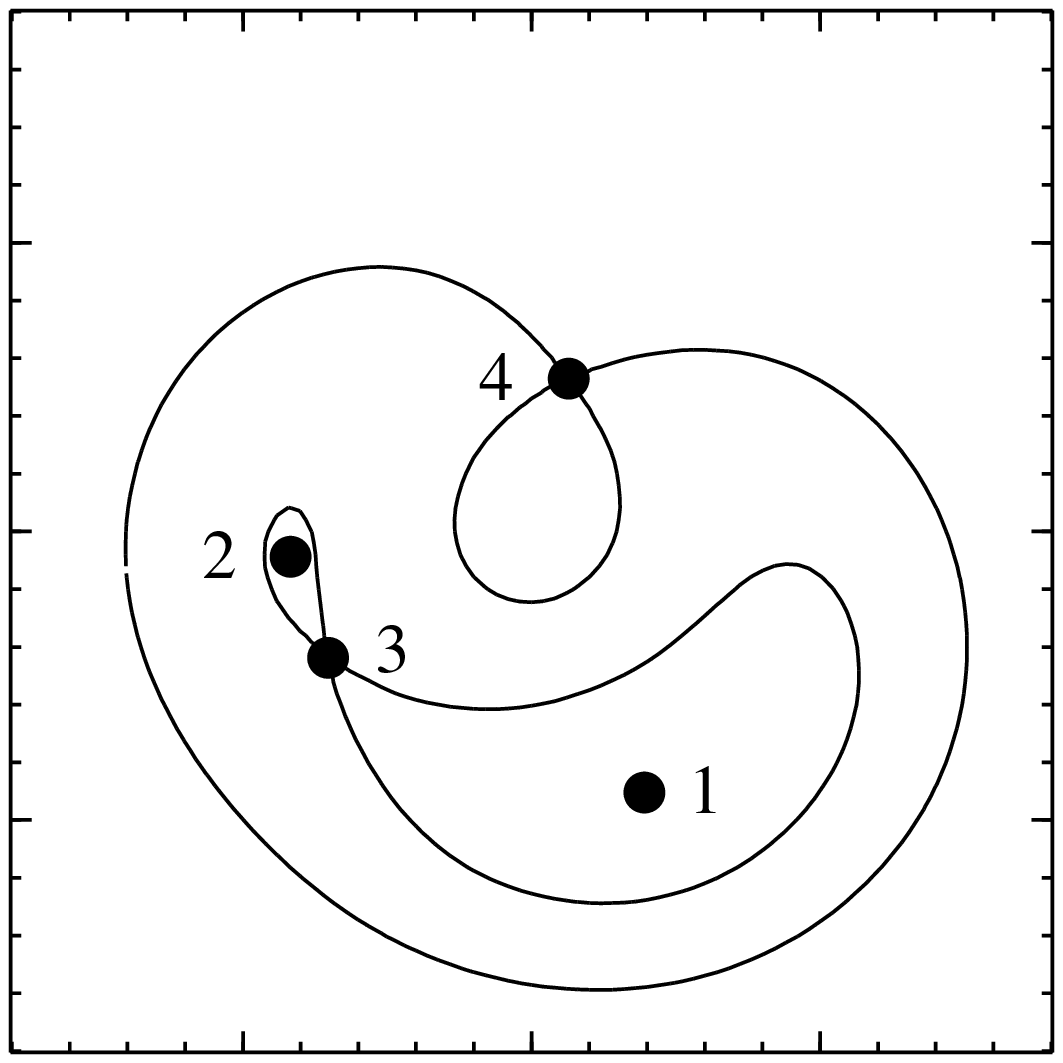}
\caption{Expected image morphology and time-ordering of the images,
assuming the center of the lens is coincident with the center of the
dominant galaxy.}
\label{arriv-a}
\end{figure}

Before making any actual models, it is helpful to use lensing theory to
identify qualitative features of the lens that are model independent. As
argued in \cite{sw03}, examining the lens morphology already provides
useful information.  In Figure \ref{arriv-a} we show the inferred
configuration of the saddle-point contours for \giant, assuming the lens
is centered on the dominant galaxy, called G1 in \cite{oguri04}.
The images are labeled according
their inferred time order.  [A different configuration with the reverse
time-ordering is also possible, but very unlikely. We will consider it
in section~\ref{timedelays}.]  As evident from the contours, images 1
and 2 are minima while 3 and 4 are saddle-points.  Since time delays
scale with the area of the lens \citep{s04} the expected time delays
would be $\sim100$ times longer than in \pg.  But apart from its giant
size the lens appears to be a typical inclined quad. The long axis would
be roughly N-S and the source somewhat SW of the lens center.

To reconstruct the lens we use the {\em PixeLens\/} code \citep{sw04}.
The input to {\em PixeLens\/} are (i)~the lensing data, which for
\giant\ are the positions of images 1--4 relative to the cluster center,
and the redshifts of the lens and source,\footnote{For definiteness we
take $H_0^{-1}=15\,{\rm Gyr}$ (or $H_0\simeq65$ in local units),
$\Omega_m=0.3,\Omega_\Lambda=0.7$ in this paper. It is necessary to fix
$H_0$ `by hypothesis' because in the absence of time-delay observations
there is nothing else to set an overall time scale in the problem.}  and
(ii)~a chosen prior, which we explain below.  The output is an ensemble
of pixelated mass maps, which we then preprocess in various ways. Each
model in the ensemble reproduces the lens data precisely, and so does
any normalized superposition of models in the ensemble.  Since
superposition of many individual maps reinforces common features of mass
maps, and diminishes random `one-time' features, the ensemble-average
map can be interpreted as a smoothed map of the inner portion of the
cluster.  Uncertainties are readily derived from the statistics of the
ensemble.  Each ensemble presented in this paper has 500 models.

Earlier implementations of free-form lens reconstruction have been
applied to the famous lensing clusters A370 and A2218
\citep{asw98a,asw98b} and the galaxy lenses \pg, B1608+656, and
B1422+231 \citep{sw97,ws00,rsw03}. {\it PixeLens\/} itself was developed
for galaxy lenses, but it can be used without change for \giant.  In
fact, because in \giant\ there are a number of cluster galaxies in the
vicinity of the
images, making the mass distribution lumpy, free-form modeling is even
more desirable than in galaxy lenses.  Free-form modeling
also has the advantage of not requiring that an artificial (and
unphysical) distinction be made between individual galaxies and cluster
halo, {\em before\/} modeling.

If only the primary constraints from lensing are used, the ensemble will
be dominated by mass maps that are highly irregular and have spurious
extra images.  Restricting the ensemble to lenses that could plausibly
be galaxies or clusters is achieved by secondary constraints, or a
prior.  {\em PixeLens\/} priors have six kinds of constraints, as
follows.

\begin{enumerate}

\item Mass pixels are non-negative: $\kappa\geq0$.\footnote{$\kappa$ is
the projected surface mass density in the lens, normalized by the critical
surface mass density for lensing.}

\item Inversion symmetry about the lens center can be imposed as an
option. Since \giant\ appears to be an asymmetric lens we will not apply
this constraint.

\item The lens must be centrally concentrated. {\em PixeLens\/}
implements this by restricting the direction of the local density
gradient, $\grad$.  The default is $\grad\leq45^\circ$, meaning that
the density gradient must point within $45^\circ$ degrees of the
center.  A very small allowed range such as $\grad\leq8^\circ$ forces
the average mass distribution to be more nearly circular. A very large
allowed range like $\grad\leq80^\circ$ tends to result in mass
``fingers'' emanating from the lens center, making the mass map look
nothing like a galaxy or cluster. 

\item The $\kappa$ of any pixel can be at most twice the average of its
neighbors: $\kappa\leq2\,\knbr$, ensuring smoothness of mass maps. (But the 
central pixel is not constrained in this way, to allow for a density cusp.) 
This is the default setting. Weakening or removing this constraint 
altogether will allow for more substructure.

\item The slope of the radial mass profile, or $\radinx$ in
$r^{-\radinx}$ is bounded.  The default is $\radinx\geq0.5$.
Note that $\radinx$ is defined here for the projected mass profile;
hence an isothermal has $\radinx=1$.

\item Finally, {\em PixeLens\/} has an optional constant external shear,
and the prior confines its allowed directions, but not its magnitude.  
We denote the shear
direction by $\shear$. For $\shear=0^\circ$ the shear will have PA along
the north-south axis, and because of the way shear is defined in lensing
theory that corresponds to external mass along the {\em east-west\/}
direction.

\end{enumerate}

In the following, we present results using four different priors.
\begin{eqnarray}
\nonumber
\hbox{Prior A1:} & & \grad \leq 45^\circ, \quad 0.25\leq\radinx\leq3, \quad
\shear=70^\circ\pm45^\circ. \\
\nonumber
\hbox{Prior A2:} & & \grad \leq 45^\circ, \quad 0.25\leq\radinx\leq3, \quad
\shear=10^\circ\pm45^\circ. \\
\nonumber
\hbox{Prior B1:} & & \grad \leq 8^\circ, \quad 0.25\leq\radinx\leq3, \quad
\shear=70^\circ\pm45^\circ. \\
\nonumber
\hbox{Prior B2:} & & \grad \leq 8^\circ, \quad 0.25\leq\radinx\leq3, \quad
\shear=10^\circ\pm45^\circ.
\end{eqnarray}
All the above priors are somewhat different from the default {\em
PixeLens\/} settings. The lower bound on the steepness constraint,
$\radinx$ has been relaxed, since the central region of the cluster
can be shallower than $\radinx=0.5$; profiles steeper than 3 are
unrealistic, and have been excluded.  The A and B-type priors differ
in their gradient constraints, $\grad$. In this paper we are
particularly interested in recovering mass associated with cluster
galaxies, and other possible substructures in the lens. These
substructures are just the deviations from the smooth circularly
symmetric mass profile (see Section~\ref{substruc}).  Therefore, two
of our priors (B1 and B2) were chosen to have a small range of allowed
$\grad$ angles.  Priors A1 and A2 adopt the default setting for
$\grad$.  The smoothness constraint has been removed altogether from
both types of priors\footnote{We also ran models with the smoothness 
constraint included; the basic results are similar to the ones presented
here.}.  This was done to facilitate the recovery of
substructure: a localized mass lump can, in principle, be more
pronounced than the default smoothness constraint allows for.

Dropping the smoothness constraint 
produces a lot of pixel-to-pixel fluctuations, which do not completely
die down even if 500 individual maps are co-added. To reduce these
fluctuations we smooth the final ensemble-average map with a filter that
spreads part of each pixel onto its eight neighbors with weights
corresponding to a Gaussian with $\sigma=\sqrt{0.5}$~pixel, or $\approx0.67''$.

The external shear directions, $\shear$, were set as follows. A1 and B1 
require the external shear to correspond roughly to the local alignment 
of galaxies, those within about $60''$ of the lens (see Fig.~13a of \oguri), 
while A2 and B2 have shear aligned similar to the typical shear direction 
in the models by \oguri. If A2/B2 are correct the shear would arise well 
outside of the central $\sim 30''$--$60''$ of the lens.

\section{Maps of the total mass}\label{massmaps}

\begin{figure}
\epsscale{0.95}
\plotone{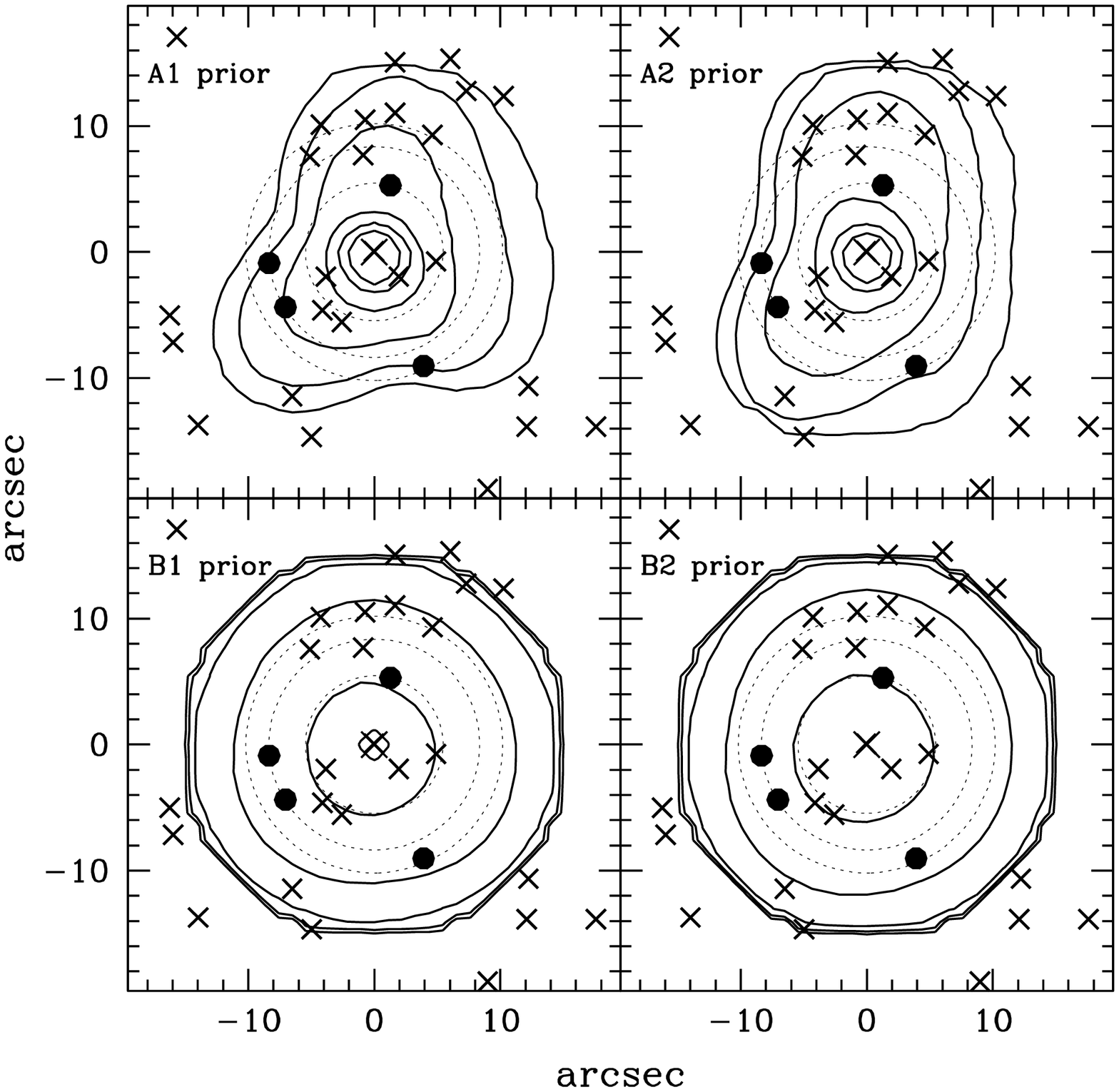}
\caption{Mass maps using different priors as indicated. Contours are
$\kappa=0.1,0.2,0.4,\ldots,3.2$ in logarithmic steps. The reconstruction
window has radius $15.2''$, the sky scale being $\simeq7.5\rm\,kpc/arcsec$.
In each panel the mass maps shown are averages from ensembles of 500;
additional smoothing has been applied to the maps using $\sigma=0.67''$
(see Section~\ref{modeling}). 
Crosses are galaxies with $i<24$, taken from Fig.~13a of \oguri. Thin
dotted circles are cluster-centered circles drawn through the images, to
guide the eye. }
\label{massmap}
\end{figure}

The four panels of Figure~\ref{massmap} show reconstructions using the
priors A1, A2, B1 and B2.  The solid dots are the QSO images, while
crosses mark the locations of cluster galaxies.

Let us consider first the top panels in Fig.~\ref{massmap} (i.e., A1 and
A2 priors).  Evidently, pre-specifying the approximate direction of
external shear affects the shape of the mass distribution to some
degree; the most notable difference between right and left panels is the
extent of the SW elongation of isodensity contours. The presence of a mass 
component in the W part of the map is to some degree degenerate with 
external shear direction $\shear=10^\circ$. This probably explains why 
the SW extension is rather pronounced in A1 ($\shear=70^\circ$) and almost 
absent in A2 ($\shear=10^\circ$). 

More interestingly,
we see that even though we did not tell {\em PixeLens\/} anything about
individual galaxies, their presence is clearly reflected in the
reconstructed mass distributions. (Crosses in the figure represent
locations of galaxies.) The density contours encompass the northern 
grouping of galaxies, and also indicate the presence of mass just south east 
of the lens center. There are several galaxies there, inside as well as
outside the image circle. The mass distribution in the very central
region of the cluster, within the radius of the innermost image, is
probably not well constrained and the two galaxies just west of the
central one are not noticeable. The slight extension of the density
contours towards SW are probably due to the 2--3 galaxies located in
that area, just beyond the image circle.

The maps in the bottom panels (priors B1 and B2) also indicate the
presence of the galaxy groupings in the N and SE parts of the map,
however, here the distortions of the isodensity contours are very
slight: one can tell that the contours are not entirely circular by
comparing them to the dotted thin circles drawn through the locations
of the images, to guide the eye. The near circular nature of the
isodensity contours is a direct consequence of restricting the range
of the direction of density gradient to a narrow cone 
($\grad \leq 8^\circ$). In Section ~\ref{substruc} we will
discuss, in more detail, how well B1 and B2 maps recover the
presence of galaxies in the cluster.

\section{Gradient of the mass profile}\label{gradient}

The mass distribution in dark matter halos is of great importance in
determining the nature of dark matter. In the central regions of clusters,
$\sim50$--$100\kpc$, the baryons are less important than in the 
central regions of individual galaxies, therefore one can reasonably
assume that the best-fit density profile is due to dark matter alone. 
In fact, our analysis in Section~\ref{substruc} will allow us to 
test this assumption, at least partially: we will estimate the fraction
of mass associated with individual cluster galaxies, but we cannot say
anything about hot diffuse smoothly distributed gas that might be 
present in the cluster.

\begin{figure}
\epsscale{0.95}
\plotone{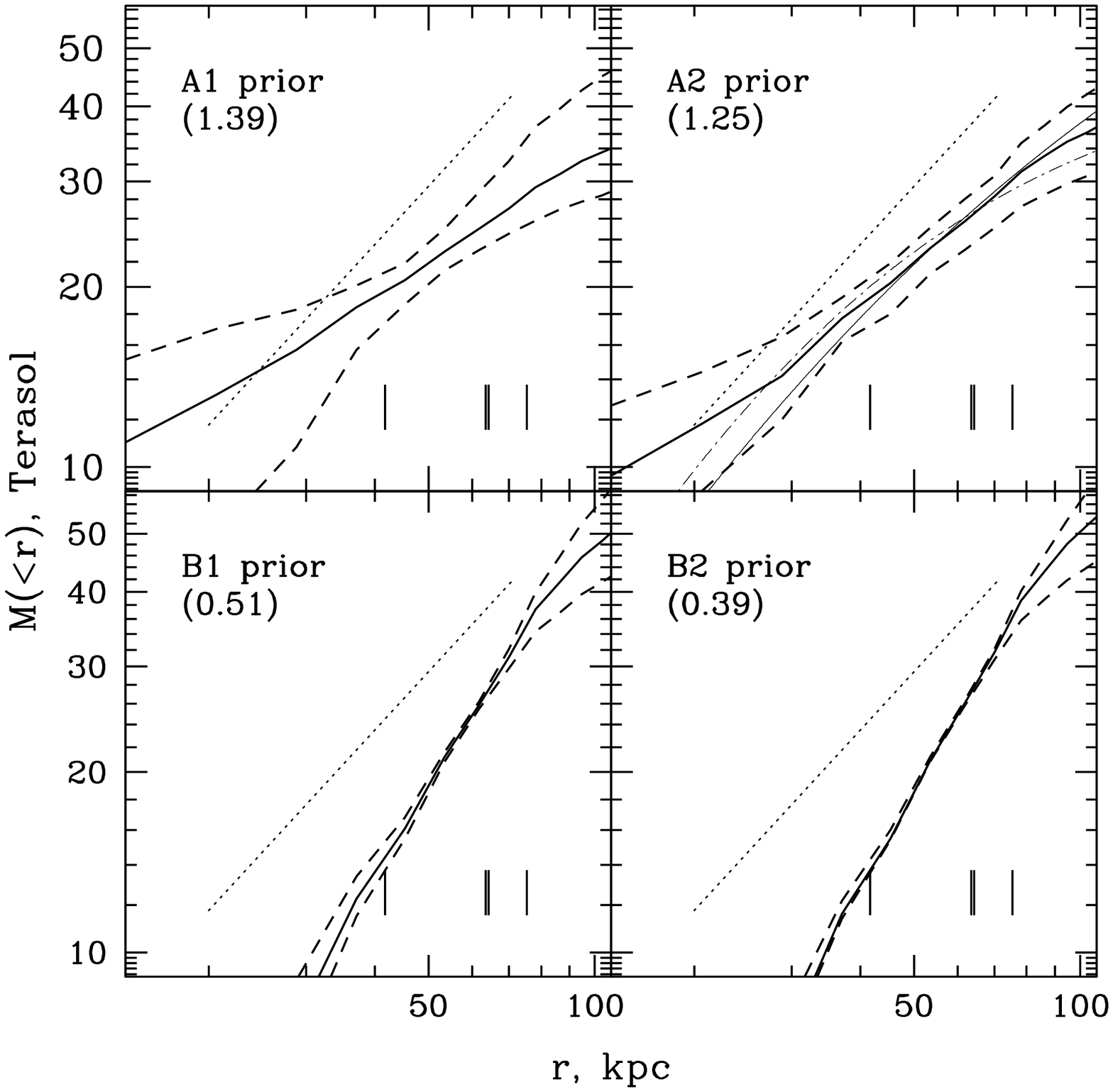}
\caption{Enclosed mass $M(<r)$ in terasol (or $10^{12}M_\odot$).  In
each panel, the thick solid line is the ensemble-median of $M(<r)$, with
the dashed lines enclosing 90\% of the individual models in the
ensemble.  The four short vertical bars at the bottom mark the
image radii; the images 2 and 3 being nearly equidistant from the lens
center, their positions are almost indistinguishable in this plot. The
number in parenthesis is the projected double logarithmic density slope
$\radinx$, while the dotted line represents the isothermal slope,
$\radinx=1$. The two thin lines in the upper right panel are some 
specific NFW (solid) and CORE (dot-dash) fits, see 
Section~\ref{gradient} for details.}
\label{enclmass}
\end{figure}

From our free-form mass maps, we derive the mass enclosed within a given
radius, $M(<r)$.  The four panels of Figure~\ref{enclmass} show $M(<r)$
for the four cases shown in Fig.~\ref{massmap}; dashed lines are 90\%
confidence limits obtained using the whole ensemble of individual
recovered mass maps. (Note that the uncertainties at different $r$ are
not independent.)  As is typical of quads, the total mass within the
image circle is very well constrained, and the errors are smallest where
$\langle\kappa(<r)\rangle=1$. But because of the mass-disk degeneracy
\citep{falco85,s00}
the distribution of the mass can take on various forms.  We expect that
the envelope delineated by the confidence limits would look like two
curves crossing at $\langle\kappa(<r)\rangle=1$, where the two
intersecting curves of the envelope represent the shallowest and the
steepest mass profiles possible. This is borne out in the plots.

Even though the density slopes from A-type and B-type priors are very
different the mass enclosed within $\sim 60\kpc$, which is the average
radius of the image annulus, is nearly identical in all four cases,
emphasizing again the robustness of enclosed mass estimate.

The thin dotted line in each panel of Fig.~\ref{enclmass} corresponds to
the isothermal slope, or $\radinx=1$. The actual value of $\radinx$ in
the image annulus is shown in parenthesis.  Comparing panels we see that
the derived slope depends strongly on the secondary constraint assumptions 
(whether an A-type or B-type prior was used) but A1/A2 and B1/B2 give
similar slopes. 

Since A-type and B-type priors result in such different density slopes
for \giant, is the derived density slope simply a consequence of the
prior, with the true density slope unrecoverable?  Comparison with the
galaxy lens \pg\ indicates that the situation is not that dire.  While
A-type and B-type priors in the case of \giant\ result in
$\radinx\approx 1.3$ and $0.4$, respectively, the same set of priors
when applied to \pg\ (but using $\shear=-45^\circ\pm45^\circ$
appropriate for this lens) result in $\radinx\approx 1.6$ and $1.2$.
Comparing the two sets of numbers suggests that the galaxy in \pg\ has
a steeper density profile than does the cluster in \giant.  To test
this hypothesis we apply another prior, with a narrow range of shallow
density slopes: $0.3\leq\radinx\leq0.5$.  If $\grad\leq 8^\circ$ {\em
PixeLens\/} finds no mass models that satisfy this set of constraints
for \pg, and if $\grad\leq 45^\circ$ ensemble mass maps have spurious
images. It seems that shallow slopes are incompatible with \pg. For
\giant\ {\em PixeLens\/} has no problem generating realistic mass maps
using shallow density profiles, and either type of $\grad$
constraints.  We conclude that the density gradient in \pg\ is around
$1$--$2$, while \giant\ is shallower, possibly as shallow as
$\radinx\approx 0.3$.  It may be possible to improve these constraints
by testing against a large number of synthetic models, analogous to
the blind tests in \cite{ws00} but much more detailed; we leave that
for the future.

It is interesting to note that B-type priors result in much better
constrained $M(<r)$ distributions (Fig.~\ref{enclmass}) than A-type
priors. This is probably due to steeper density profiles (as result from 
A-type priors) being more vulnerable to mass disk degeneracy, and conversely, 
shallower profiles resulting from B-type priors leaving little room for 
mass disk degeneracy.

Since $M(<r)$ is a circularly averaged quantity, it can be used to
constrain `average' halo density profile models.  To do this we fit two
types of density profiles: (i)~$\rho\propto(1+r/r_c)^{-\gamma}$ having a
flat density core (called `CORE' models), and (ii)~NFW ~\citep{nfw97}, with
$\rho\propto [(r/r_s)(1+r/r_s)^2]^{-1}$. In the latter case, scale radius,
$r_s$, and virial radius, $r_{200}$, are related through the
concentration parameter, $c=r_{200}/r_s$, and take on values of about
4--5 for rich galaxy clusters.

\begin{figure}
\epsscale{0.9}
\plotone{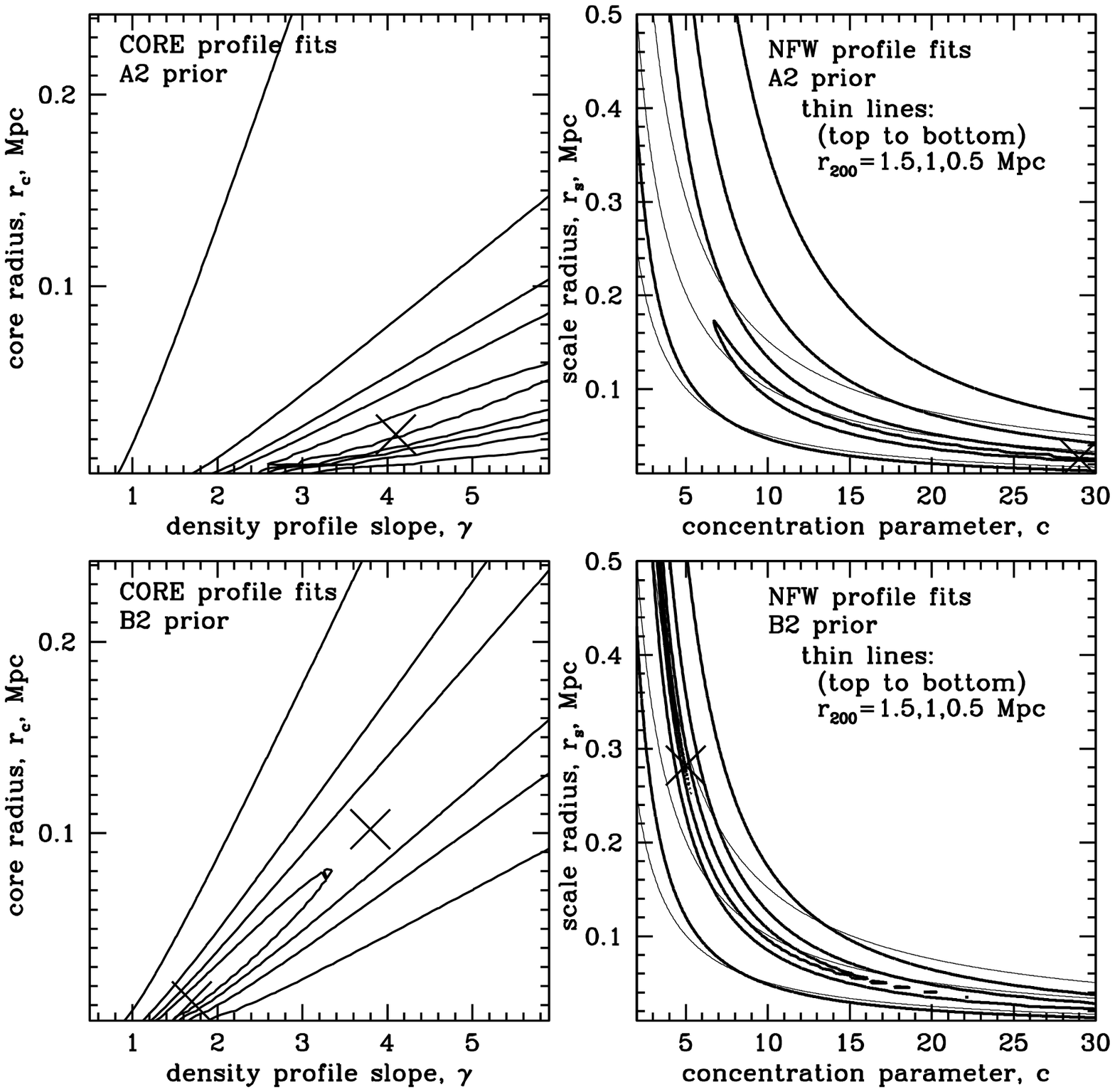}
\caption{Simple parametric fits to free-form mass maps generated using
A2 (top panels) and B2 (bottom panels) priors.  The fits are derived via 
the enclosed mass, and solid lines are contours of constant $\chi^2$. 
Left panels show CORE models of the form $\rho\propto(1+r/r_c)^{-\gamma}$;
$\chi^2$ contours are at 1, 3, 5, 10, 30 per dof. 
Right panels show NFW model fits where concentration parameter and scale 
radius were allowed to vary independently; $\chi^2$ contours are at 
10, 100, 1000, 10000 per dof. The $\chi^2$ increase very rapidly away 
from the locus of best-fit NFW models; for example, the model marked with 
a cross in the bottom right panel has a $\chi^2\approx 1.5$. (The `islands' 
are artifacts of the grid and should be disregarded.) The crosses 
in all four panels show models mentioned in Section~\ref{gradient}. 
The thin lines in the two right panels are curves of constant virial 
radius, $r_{200}$ (or, equivalently constant halo mass); 
$r_{200}=1.5, 1, 0.5$ Mpc from top to bottom.}
\label{grid}
\end{figure}

The goodness of fit of NFW/CORE models will be ascertained using
$\chi^2$, which requires rms values.  The distribution of $M(<r)$ values
for any given $r$ for an ensemble of models is not necessarily Gaussian,
and is often asymmetric, but for the purposes of estimating rms of the
distribution we assume that the size of the 90\% error-bars (which we 
obtain from {\em PixeLens\/}) is 1.65
times the size of the 68\% error-bars, the latter being $1\sigma$ for a
Gaussian. We use these estimates of $\sigma$ to calculate $\chi^2$ for
the model fits. The $\chi^2$ contour levels using A2 and B2 mass maps are 
shown in Fig.~\ref{grid}, but A1 and B1 maps would have produced similar 
results. The plotted $\chi^2$ contour levels are indicated in the figure 
caption, however, these values are not to be taken too seriously because our
assumption of symmetric, Gaussian distributed errors is not always very
good. The non-smoothness of some contours in these plots is partly due
to that. 

The top panels of Fig.~\ref{grid} show the $\chi^2$ contours when 
CORE and NFW density models are fitted to the enclosed mass data of
A2 prior models. These density distributions are rather steep, so
neither CORE nor NFW provide very good fits. Some of the better
CORE and NFW fits
(marked with a cross in each of the two top panels) are plotted
in the top right panel of Fig.~\ref{enclmass} with thin dot-dash and 
solid lines, respectively. The CORE model requires the density outside
of the central $\sim 30$~kpc to be very steep, $\alpha=4$, while NFW
fit has a concentration parameter of $c\approx30$, well outside the
typical range for galaxy clusters. Thus, A2 prior mass models
seem to be unrealistic.

The bottom panels of Fig.~\ref{grid} show the $\chi^2$ contours when 
CORE and NFW density models are fitted to the enclosed mass data of
B2 prior models. These density distributions are shallow, and could 
easily accommodate flat density cores as large as 100 kpc. The two 
crosses in the bottom left panel mark some specific models;
if the corresponding models were plotted in the bottom right panel 
of Fig.~\ref{enclmass} they would be indistinguishable from the actual 
line representing $M(<r)$; we did not overplot these. 
The thin lines in the bottom right panel (NFW fits) show contours of 
constant virial radius of NFW halos. The virial radius, and
hence the mass of the best fitting halos are relatively well constrained;
$r_{200}\approx$ 1--1.5 kpc. The halo marked with a cross provides an
excellent fit to the data. This halo's concentration parameter is $c=5$,
and its mass is $M_{200}\approx 4.17\times10^{14} M_\odot$, which is
comparable to the mass estimated by \oguri, 
~3.1--4.6$\times10^{14} h_{65}^{-1}M_\odot$. 

We conclude that our fits, using B2 prior models, to the circularly
averaged density distribution of the cluster are very similar to those
in \oguri.  But our free-form models can be taken much further than
parametric models, revealing the deviations from the average density
profile, i.e., substructure in the mass distribution.

\section{Detecting substructure}\label{substruc}

The term `substructure' encompasses a host of different types of 
deviations of the lensing mass distribution from smooth. There is
compact substructure, like stars or globular clusters in galaxies, 
and smooth substructure, like dwarf dark matter halos around 
galaxies. Compact substructure can result in micro- and/or milli-lensing, 
whereby many unresolved micro- and milli-images are produced. 
Micro-lensing and milli-lensing lead to a wealth of observable 
consequences \citep{wambs90,schech02,ws95} some of which are yet 
to be detected.

Smooth mass lumps in the main lens do not split the macro-images
into sub-images, but they can affect the fluxes, and even
the positions of macro-images. Observed flux ratios of macro-images have
been extensively used in the recent literature to look for smooth
substructure in and around galaxies \citep{dk02,mz02,chiba02,ew03}.

Here, we look for substructure using positions of macro-images only.
To our knowledge, this is the first work to do so. Also, our macro-lens
is a cluster of galaxies, not an individual galaxy as is the case in
most flux-ratio substructure investigations, hence the mass range
of substructure lumps that we find (see below) is necessarily larger 
than that searched for in individual galaxies.

To examine the substructure in the central regions of \giant, 
at every point in the map we subtract the circularly
averaged surface mass density for that radius (i.e. distance from 
cluster center), and thus obtain a residual mass map.

\begin{figure}
\epsscale{0.95}
\plotone{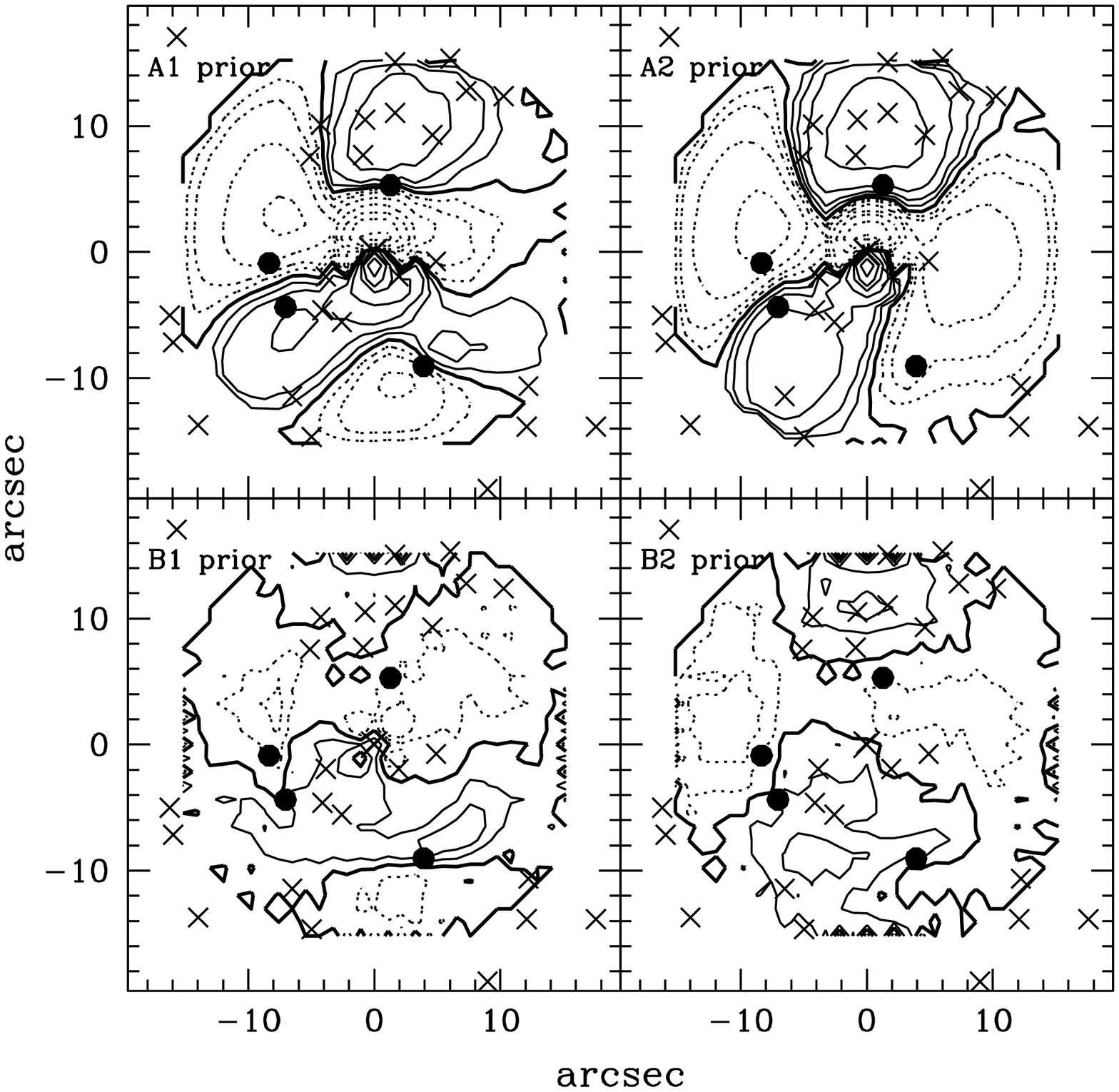}
\caption{Residuals after subtracting off the circularly averaged density
profile in each panel of Figure \ref{massmap}.  The thick contour is now
$\Delta\kappa=0$, the thin solid lines are
$\Delta\kappa=0.025, 0.05,0.1,0.2,\ldots$ and the thin dotted lines are
$\Delta\kappa=-0.025, -0.05,-0.1,-0.2,\ldots$ Crosses mark the positions
of galaxies with $i<24$.}
\label{resmap}
\end{figure}

Figure~\ref{resmap} shows the residual mass distributions corresponding
to the mass maps in Figure~\ref{massmap}.  All four reconstructions
detect the main substructure components in the cluster: the galaxy group
N of image~4, and the galaxy group SE of images~2 and 3.  The only 
galaxy which is consistently under-detected is one $\sim2''$ W of
center; this is probably because the galaxy is well inside the
region of lensing constraints, therefore it is not clear how much stock
we should put into the central features of the map.  It might at first
be surprising that given the smoothness and near circular appearance of
the iso-density contours in the bottom panels of Fig~\ref{massmap},
the substructure in the bottom panels of Fig.~\ref{resmap} is very well
delineated. The explanation lies in the shallow average density profile
of B-type prior maps: it is easy to `hide' substructure superimposed 
on a nearly flat density field.

The most obvious difference between A1 and A2, and similarly between
B1 and B2 is the presence of a mass lump in the SW part of the 
A1 and B1 maps. This mass component does not correspond to any visible 
galaxies. The simplest explanation is that A1 and B1 assume the wrong 
external shear direction ($\shear=70^\circ\pm45^\circ$), while the shear 
direction incorporated into the A2 and B2 priors ($\shear=10^\circ\pm45^\circ$) 
is close to the true one. The latter is the same shear direction as
obtained in \oguri. A2 and B2 maps have no extended mass features
that do not correspond to visible galaxies. From now on we will concentrate 
mainly on the A2 and B2 prior models.

Having now a good visual match to the galaxies in the residual mass maps, 
we test the substructure more quantitatively. Let us 
define $M_l(\btheta)$ as the residual mass within $l$ pixels of a point
$\btheta$ of the residual mass map A2 or B2.  Then a choice of 
points $\{\bthetan\}$
will result in a distribution $\{M_l(\bthetan)\}$.  If the mass
residuals are uncorrelated with the galaxies (the null hypothesis), 
then the distribution of $\{M_l(\bthetan)\}$, where $\bthetan$ are galaxy 
positions should not be significantly different from the $\{M_l(\bthetan)\}$ 
distribution obtained using random $\bthetan$ positions.  
Using the 19 galaxies with $i<24$ taken from \oguri's  Fig.~13a, 
we find that a Kolmogorov-Smirnov test rejects the null hypothesis at
significance levels of 99.92\%, 99.91\%, 99.75\%, 95.8\% confidence
levels, for $l=2,4,6,8$ respectively, for A2 prior model.
For B2, the corresponding confidence levels are 
99.28\%, 99.37\%, 97.2\%, 98.94\%. Thus, we unambiguously detect 
mass associated with galaxies in cluster \giant.

It should also be possible to infer the typical size of the galaxies, 
but because the galaxies are heavily clustered, especially in the N part 
of the cluster, we can only estimate an approximate upper limit on size. 
We define upper limit as the size corresponding to the $l$ value above 
which the enclosed residual mass summed over all galaxies begins to 
decrease. This corresponds to 4 pixels, or $\sim30\kpc$, for both A2 and B2.

We can also estimate the mass associated with the cluster galaxies, by adding
up the residual mass of all pixels that lie within $l$ pixels of all 19 
galaxies and then multiplying by 2 as a rough correction for the fact
that residual mass can be negative or positive.  For $l=1$ and $l=4$,
the average mass per galaxy is 
$1.2\times 10^{11}\,M_\odot$ and $1.9\times 10^{11}\,M_\odot$ 
respectively, for A2, and
$2.1\times 10^{10}\,M_\odot$ and $5.1\times 10^{10}\,M_\odot$ for B2
prior model. For a 64-fold change in
galaxy volume, there is only a $\sim 2$-fold change in the derived average
galaxy mass, for both priors.  This small change must be partly due to the 
galaxies being heavily clustered, but it also indicates that the derived 
mass is quite robust.

From the photometry presented in \oguri\ (see their Fig. 10) we
estimate that a typical galaxy (of the 19 closest to the images), has
$M_r\sim -20.5$, or about $1.6\times 10^{10}~L_\odot$. Taking a
typical galaxy mass to correspond to the mass enclosed within the 4
pixels, we obtain M/L ratios of 11.2 and 3.2, for A2 and B2
respectively.  This is compatible with the galaxies in the center of
the cluster being composed entirely of stars. Since we know the total
enclosed mass in the reconstructed region, and the total mass of the
galaxies, we can estimate the fraction of mass contained in the
galaxies; this comes to about 9\%, and 2\% for A2 and B2
respectively.  This is also the lower limit on the baryonic content of
the inner $100\kpc$ of the cluster; we have no means of determining
the mass in hot diffuse smoothly distributed gas in the cluster.

The values of M/L ratios and mass fraction in galaxies derived using
priors A2 and B2 can be interpreted as defining the range of
uncertainty in these quantities.  We found further {\em PixeLens\/}
reconstructions using different sets of priors to be generally
consistent with this interpretation. 

If \giant\ is indeed a scaled-up version of a galaxy lens like \pg,
the implications are interesting.  If the latter system has $\sim$2--10\%
of its mass in the same type of substructure as \giant, we would
expect to detect a total mass of $(0.6-3)\times 10^{10}\,M_\odot$
distributed in several individual lumps of $\sim 10^9\,M_\odot$.
Although we would not detect individual lumps, their expected mass
range is typical of dwarf dark matter halos currently being searched
for around individual galaxies.

\begin{figure}
\epsscale{0.95}
\plotone{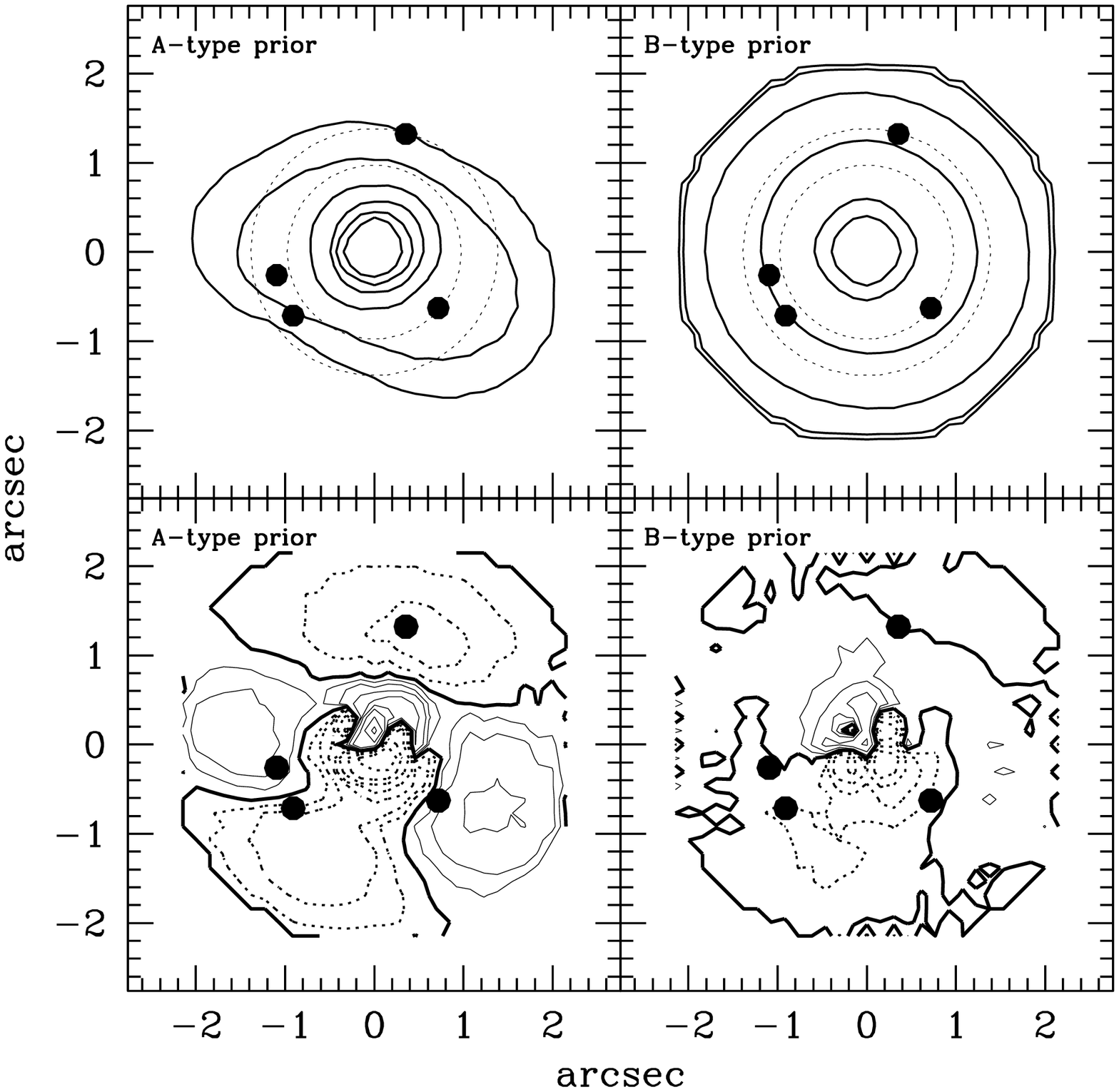}
\caption{Results from models of \pg\ using A (left panels) and 
B-type (right panels) priors, and $\shear=-45^\circ\pm45^\circ$.
{\em Upper:\/} mass maps analogous to Figure~\ref{massmap}.  
{\em Lower:\/} residual mass maps analogous to Figure~\ref{resmap}.}
\label{map1115}
\end{figure}

Accordingly, we reconstruct \pg\ in a way as similar as possible to our
reconstruction of \giant: we use an A and B-type priors
(with $\shear=-45^\circ\pm45^\circ$ appropriate for \pg), allow
the lens to be asymmetric, and disregard the observed time delays,
supplying {\em PixeLens} only the image positions.  The resulting maps
are shown in Fig.~\ref{map1115}.  It is clear that the mass
distribution in \pg\ is considerably smoother than that of the
corresponding maps of \giant, except for the central region. Also, 
the map is very nearly inversion-symmetric, unlike \giant.  This 
result indicates that the halo of \pg\ is not populated with numerous 
dwarf dark matter halos. Comparison of the residual maps
also removes any remaining concerns that our detection of
substructure in \giant\ could be an artifact.

\section{Time delay predictions}\label{timedelays}

In a model-ensemble generated by {\em PixeLens,} each model has a set of
time delays.  An ensemble of time delays is conveniently plotted as a
histogram and immediately provides predicted time delays and
uncertainties.

Nearly all the models in \oguri\ have the time-ordering shown in
Figure~\ref{arriv-a}, but a few have the reverse ordering. We can
reproduce the reverse time-order, and the appropriately modified version
of Figure~\ref{arriv-a}, if we assume the lens center is $\simeq3''$ SW
of the dominant galaxy.  (We do not know whether a similar offset
applies in the \oguri\ models, but expect it is probably the case.)  But
shifting the lens center also tends to allow a lot of models with
spurious extra images.  Hence the reverse time-order seems very
unlikely, and we will not consider it further.

\begin{figure}
\epsscale{0.4}
\plotone{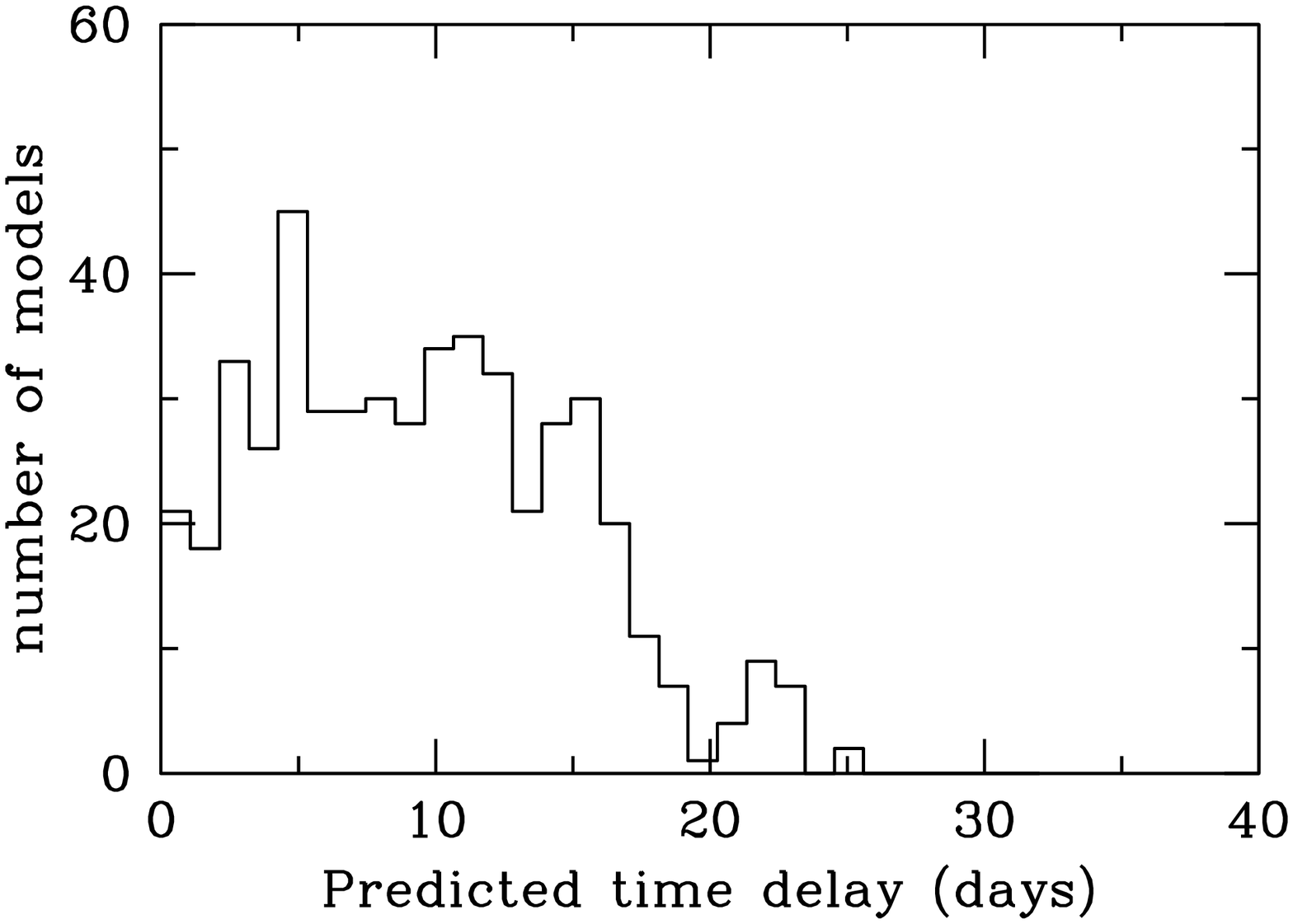}  
\plotone{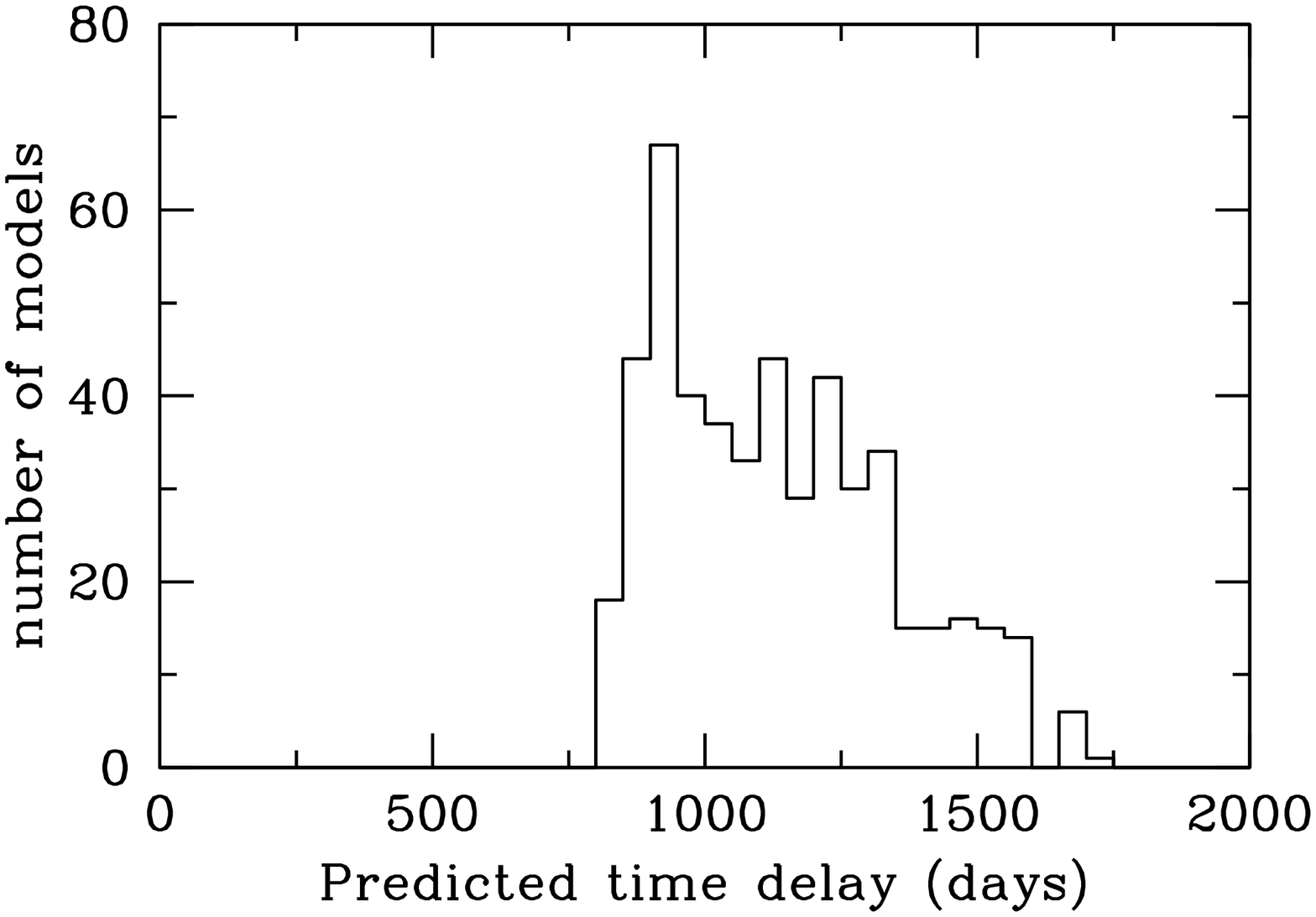}  
\plotone{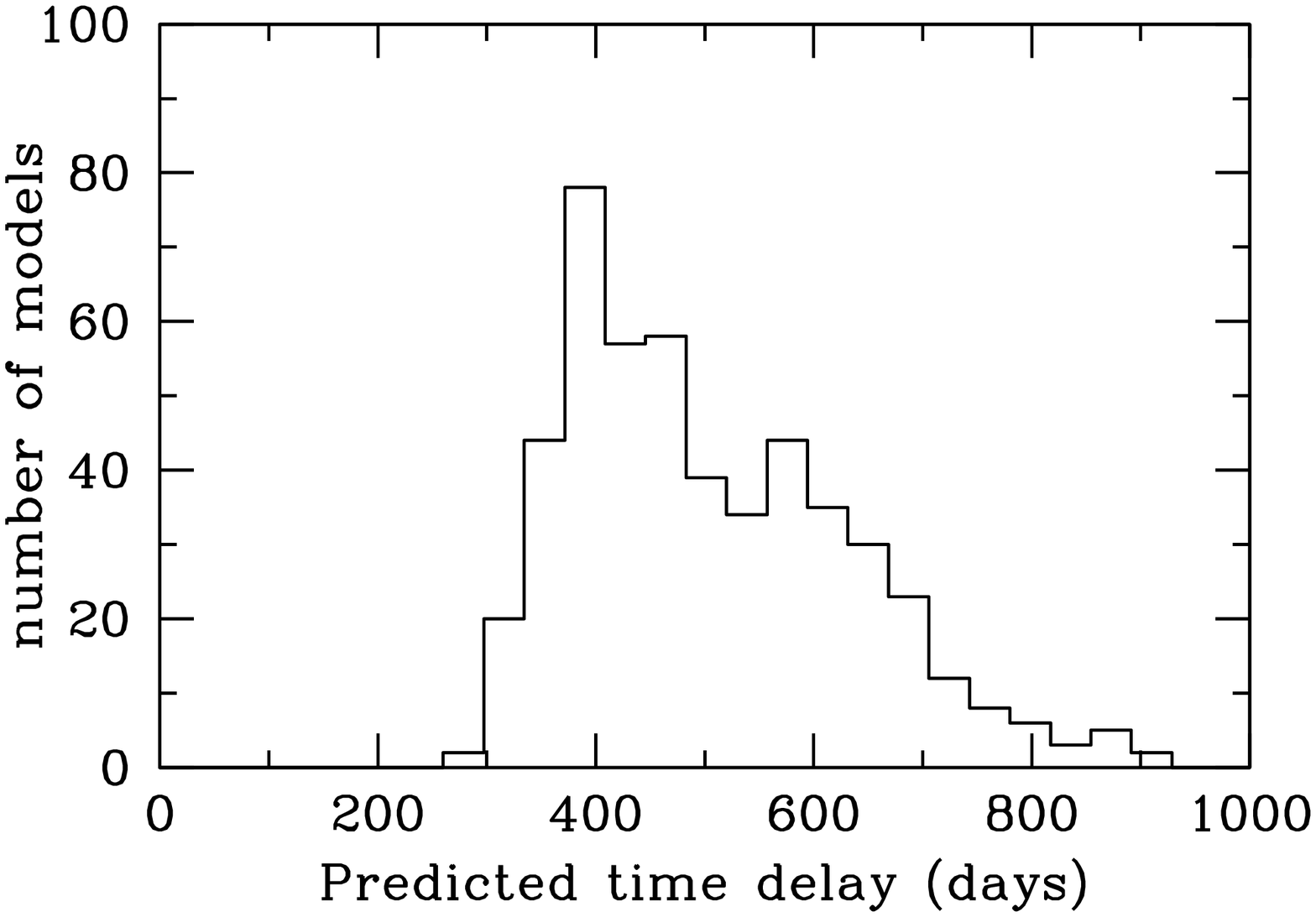}  
\plotone{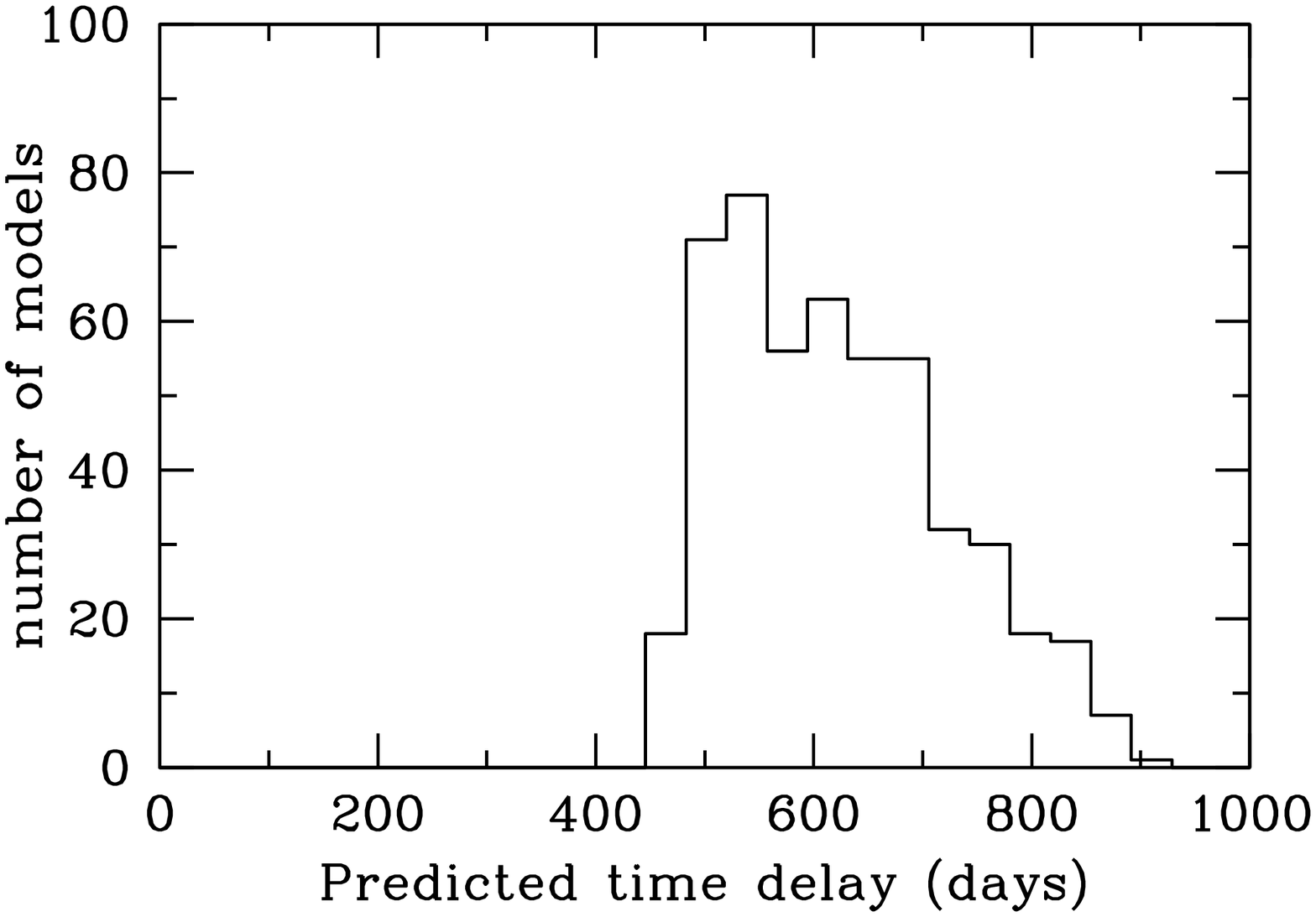}  
\caption{Predicted time delays for \giant, from reconstructions using
a B2 prior. Upper left: images 2--3, upper right: 1--4, lower left:
1--2, lower right: 3--4.}
\label{delay-a}
\end{figure}

Figure~\ref{delay-a} shows the distribution of predicted time delays
between different images for the models of the B2 prior, which give 
shallow mass profiles for the cluster, similar to those of \oguri.
The possible range of time-delays is large,
but the spread is similar to other quads---see e.g., \cite{rsw03}. Apart
from the scale, the predicted time delays are as one would expect for
galaxy lenses. The time delays for the A2 prior models (not shown) are 
about 4 times longer than those for B2, as would be expected for
the deeper potential wells of the A2 models.

\begin{figure}
\epsscale{0.45}
\plotone{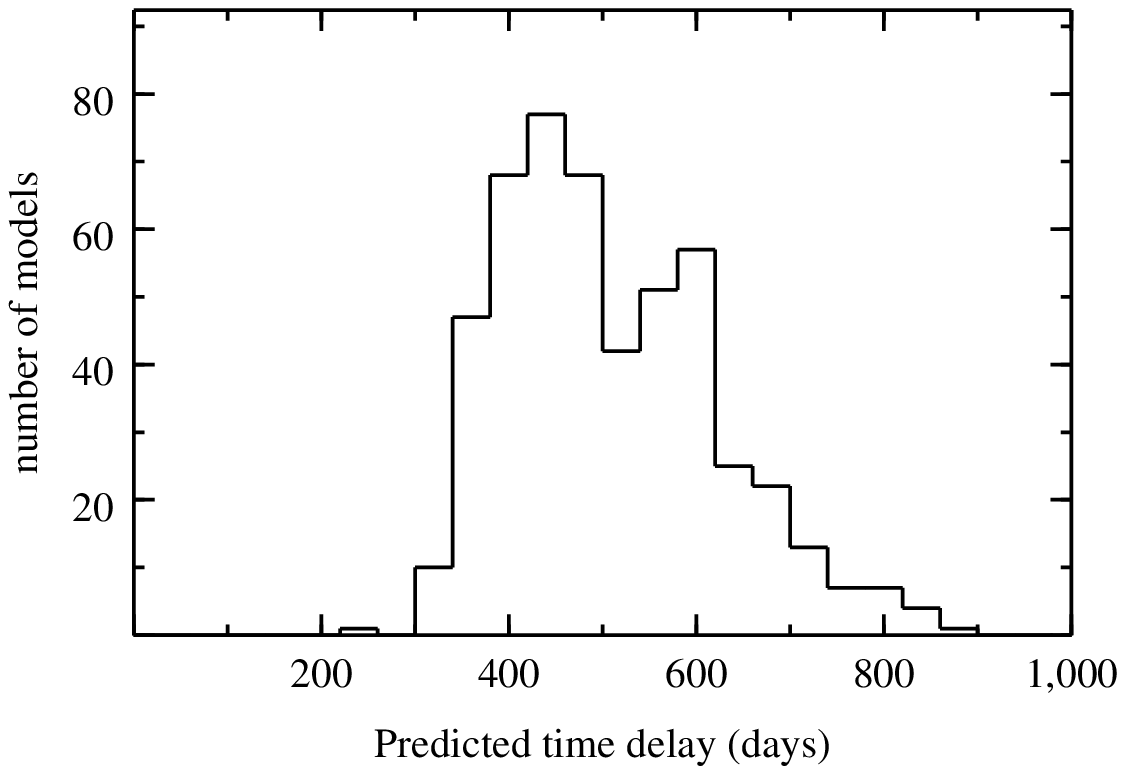}  
\plotone{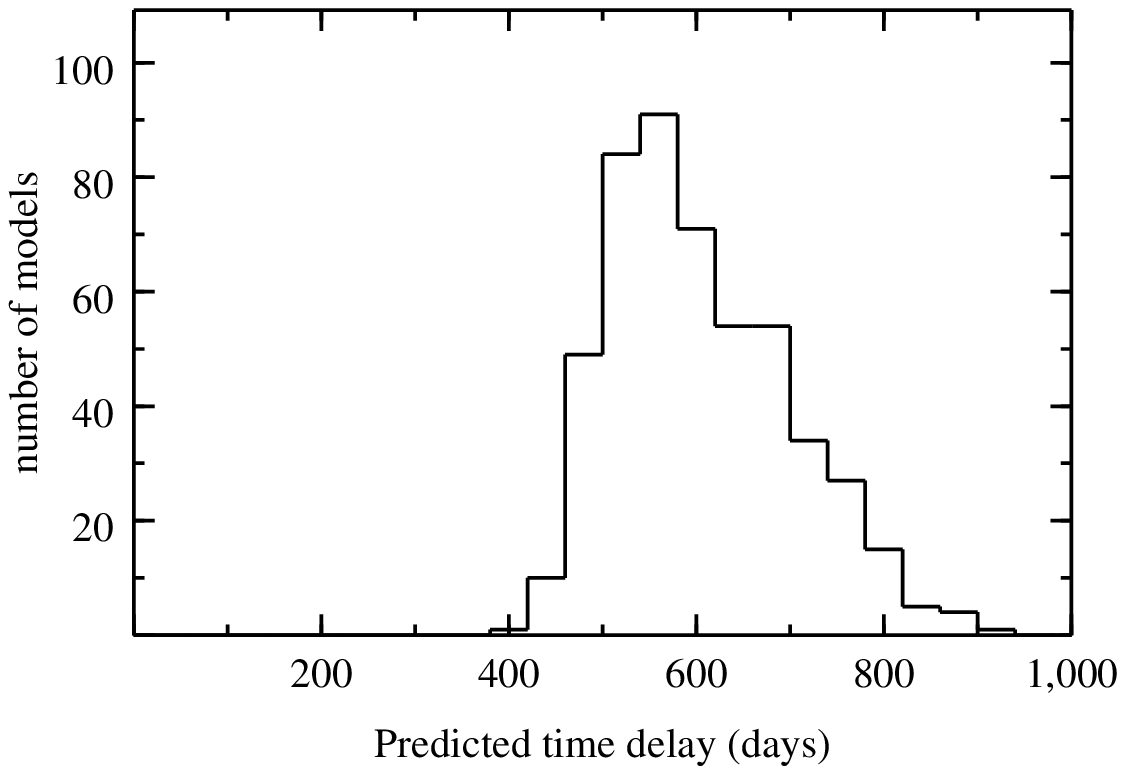}  
\caption{Like the two lower panels of Figure \ref{delay-a}, but with the
2--3 delay fixed at 10 days.}
\label{delay-b}
\end{figure}

The parametric models by \oguri\ show a similar range of
predicted time delays (see their Fig.~19), with the typical delays for
2--3 and 1--4 image pairs comparable to what we have for B-prior
reconstructions of \giant\ (upper panels of Fig.~\ref{delay-a}).
Interestingly, their models also show a strong correlation between the
2--3 and 1--4 time delays (i.e., the shortest and longest).  An easy way
to look for such a correlation using {\em PixeLens\/} is simply to fix
one of the time delays.  Accordingly, we fixed the shortest time delay
at 10 days and generated another ensemble of models.  Figure
\ref{delay-b} shows the 1--2 and 3--4 time delays with the extra
constraint.  We see that the spread gets smaller but not very much
smaller.

Comparing the time-delay results leads to an interesting point
regarding parametric versus free-form lens models.  Time delays
varying over a large range while staying in the same ratio are a
signature of the mass-disk degeneracy, which makes lens models steeper
while lengthening all time delays uniformly---see \cite{s00} for
details of this simple interpretation.  Other lensing degeneracies do
not necessarily share this property.  Thus we conclude that the
\oguri\ models are exploring the mass-disk or steepness
degeneracy very well, but are much more restricted with regard to
other degeneracies.  On the other hand, in {\em PixeLens\/} the
mass-disk degeneracy, though important, is no longer dominant.

Even though it would not strongly constrain the lens, a measurement of
the 2--3 time delay in \giant\ would be very interesting for three reasons. 
First,
it would test the inferred time-ordering.  Second, time delays between
close pairs of images are not normally measurable---in galaxy lenses
they would be less than a day---only in \giant\ the giant scale brings
the delay to a convenient length.  On the other hand, the longer time
delays, which in galaxy lenses are more practical to measure, in \giant\
would be several years and hence much more difficult to measure.  Third,
a 2--3 time-delay measurement would be a test of the relatively shallow
mass profile.  The shallow profiles from B2
priors predict a delay of $\sim 10$ days, whereas the steeper models from
A2 priors give time delays of $\sim 50$ days.

\section{A ring in \giant?}

\begin{figure}
\epsscale{0.5}
\plotone{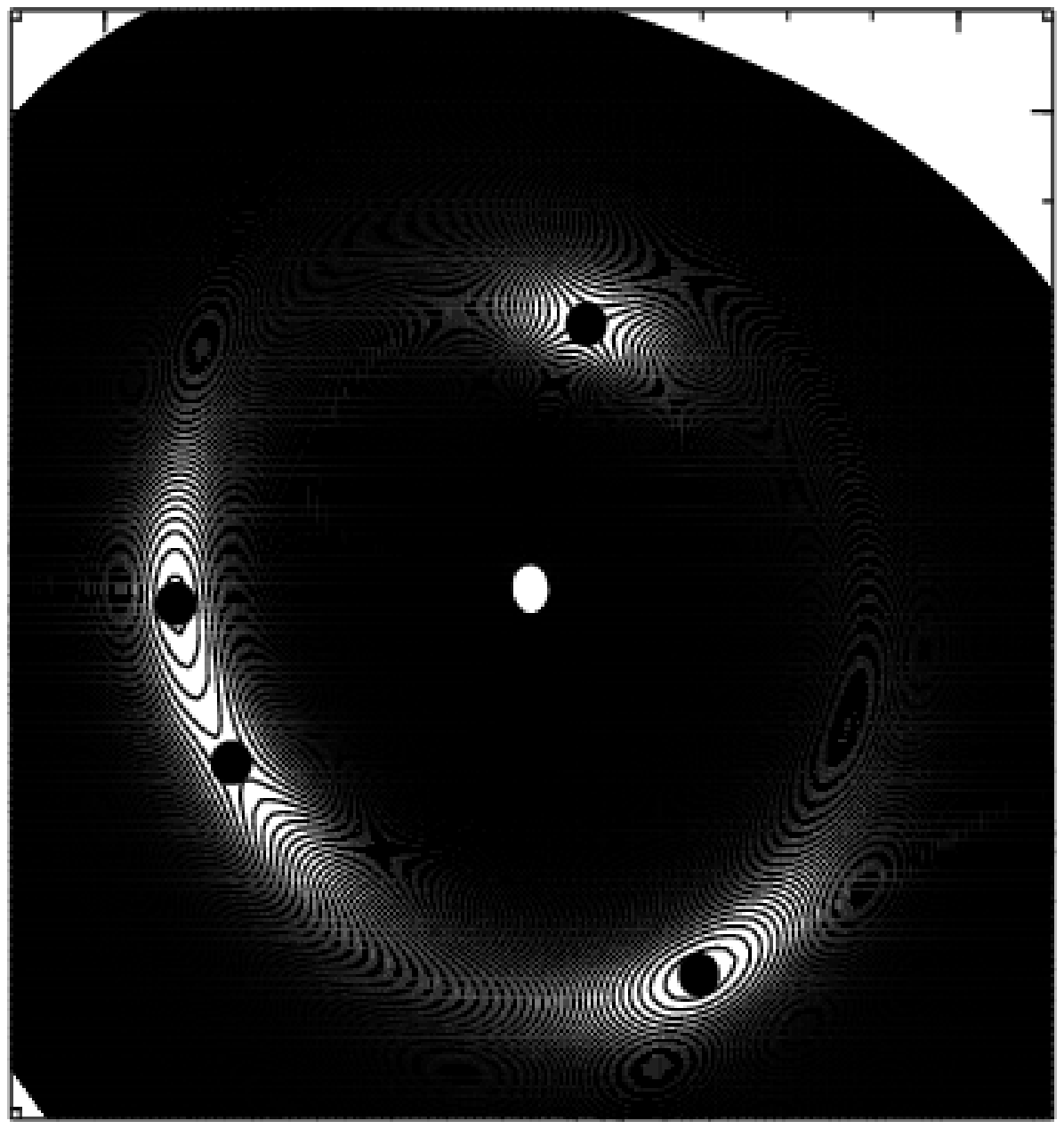}
\plotone{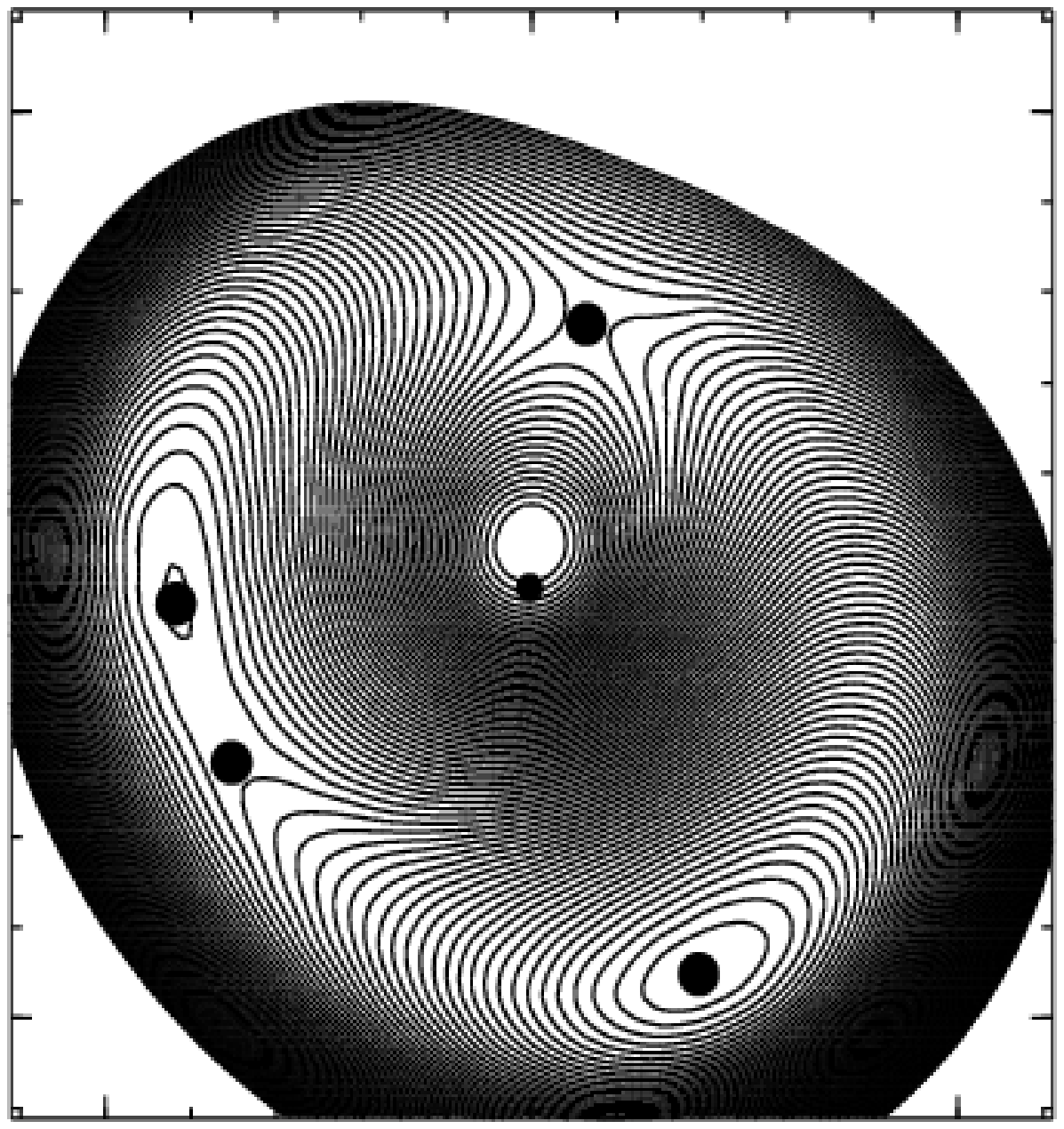}
\caption{Predicted ring, assuming the quasar host galaxy has a conical
profile of radius $\sim 2''$. 
Upper and lower panels are for A2 and B2 prior models, respectively.}
\label{ring-a}
\end{figure}

Lenses with small deviations from circular symmetry produce the best
rings; B1938+666 being the best example \citep{king98}. Asymmetric and
lumpy lenses produces short ring segments only, so \giant\ is not
expected to be much of a spectacle in that regard.  Nevertheless, a
partial ring is possible.  In Figure~\ref{ring-a} we show partial
rings such as would be produced by a source (host galaxy) with a
conical light profile of radius $2''$ with the A2 and B2 priors.  It
is really the arrival-time surface of the QSO with closely-spaced
contours, but it mimics a ring through a surprising side-effect of
lensing theory \citep{sw01}. The width of the ring depends on the
steepness of the density profile of the lens; for the same source,
the steep profile of A2 model results in a
narrow ring, while the much shallower profile of B2 gives rise to a
wider ring.

Prominent partial rings such as in those in Figure~\ref{ring-a} would 
require a large host galaxy.  That seems unlikely, probably an 
incipient ring in the form of arc-like features around the QSO images 
is more to be expected.

\section{Conclusions}

Although physically a cluster lens, \giant\ is very like a
Brobdingnagian version of a galaxy lens like \pg, with image separations
$\sim10$ larger and the expected time delays $\sim100$ times longer.
Thus the 2--3 time delay goes from hours to weeks while the 1--2 and
3--4 time delays go from weeks to years, an interesting reversal of what
is observationally most accessible.  We argue that unless the QSO host 
galaxy is rather large an Einstein ring is not expected in \giant,
but a partial ring is very likely.

The projected mass enclosed by the lens is very well constrained as
$\approx2.5\times10^{13}M_\odot$ within $60\kpc$, and
$\approx5.5\times10^{13}M_\odot$ within $100\kpc$. The circularly
averaged density profile is well fit by NFW halos, a virial mass of
$\approx4.2\times10^{14}M_\odot$ being quite robust but the
concentration parameter being uncertain. Density profiles with flat 
central cores of $r_c<100\kpc$ are not excluded.

Thus far our results are similar to those of \oguri, who used a 
combination of parametric descriptions of the central galaxy, cluster, 
and external shear to model QSO image positions. 

Our most interesting new result, however, is the detection of structure
in the lens, i.e. deviations of the mass distribution in the cluster from smooth.
We detect the structure using image positions (not flux ratios) as model input.
The structure in \giant\ is clearly associated with cluster galaxies, as confirmed
by a KS test.  A ``control'' analysis of \pg\ reveals no analogous structures. 
Although our mass maps cannot detect individual cluster galaxies (partly 
because they are so heavily clustered), we can still
estimate their mass content, which comprises about 2\%--10\% of cluster
mass. This implies typical galaxy mass-to-light ratios of about $3-11$, 
and  indicates that dark matter in the inner region of the cluster is mostly 
not bound to individual galaxies.

\end{document}